\documentclass[journal]{IEEEtran}
\usepackage{mathrsfs}
\usepackage{bbm}
\usepackage{amssymb}
\usepackage{bbding}
\usepackage{threeparttable}
\usepackage{colortbl}
\usepackage[mathcal]{euscript}

\usepackage{psfrag,calc,url,bm}
 
\usepackage{cite}

\usepackage{graphicx}

\usepackage{psfrag}

\usepackage{subfigure}
\usepackage{hyperref}
\usepackage{url}

\usepackage{stfloats}

\usepackage{amsmath}

\usepackage{float}

\usepackage{algorithmic}

\usepackage[ruled,linesnumbered,vlined]{algorithm2e}

\usepackage{color}

\usepackage{boxedminipage}

\usepackage{amsthm}

\usepackage{multirow}

\usepackage{setspace}

\usepackage{soul}
\usepackage{epstopdf}

\usepackage{svg}

\usepackage{makecell}
\usepackage{diagbox}
\allowdisplaybreaks[3]

\begin{document}

\title{NOMANet: A Graph Neural Network Enabled Power Allocation Scheme for NOMA}
\author{Yipu Hou, Yang Lu,~\IEEEmembership{Member,~IEEE}, Wei Chen~\IEEEmembership{Senior Member,~IEEE}, Bo Ai,~\IEEEmembership{Fellow,~IEEE}, \\ Dusit Niyato,~\IEEEmembership{Fellow,~IEEE}, and Zhiguo Ding,~\IEEEmembership{Fellow,~IEEE}
\thanks{Yipu Hou and Yang Lu are with the State Key Laboratory of Advanced Rail Autonomous Operation, and also with the School of Computer Science and Technology, Beijing Jiaotong University, Beijing 100044, China (e-mail: 23120351@bjtu.edu.cn, yanglu@bjtu.edu.cn).}
\thanks{Wei Chen and Bo Ai are with the School of Electronics and Information Engineering, Beijing Jiaotong University, Beijing 100044, China (e-mail: weich@bjtu.edu.cn, boai@bjtu.edu.cn).}
\thanks{Dusit Niyato is with the College of Computing and Data Science, Nanyang Technological University, Singapore 639798 (e-mail:  dniyato@ntu.edu.sg).}
\thanks{Zhiguo Ding is with Department of Electrical Engineering and Computer Science, Khalifa University, Abu Dhabi 127788, UAE (e-mail: zhiguo.ding@ieee.org).}
}

\maketitle

\begin{abstract}
This paper proposes a graph neural network (GNN) enabled power allocation scheme for non-orthogonal multiple access (NOMA) networks. In particular, a downlink scenario with one base station serving multiple users over several subchannels is considered, where the number of subchannels is less than the number of users, and thus, some users have to share a subchannel via NOMA. Our goal is to maximize the system energy efficiency subject to the rate requirement of each user and the overall budget. We  propose a deep learning based approach termed NOMA net (NOMANet) to address the considered problem. Particularly,  NOMANet is GNN-based, which maps channel state information to the desired power allocation  scheme for all subchannels. The multi-head attention and the  residual/dense connection are adopted to enhance the feature extraction. The output of  NOMANet is guaranteed to be feasible via the customized activation function and the  penalty method. Numerical results show that  NOMANet trained unsupervised achieves performance close to that of the successive convex approximation method but with a faster inference speed by about $700$ times. Besides,  NOMANet is featured by its scalability to both users and subchannels.

\end{abstract}


\section{INTRODUCTION}

The significant increase of wireless devices is encountered by the limited spectrum resource. The traditional orthogonal multiple access (OMA) technology, which assigns one orthogonal resource block to one device only, may not support the demand on high data traffic. In the past decade, non-orthogonal multiple access (NOMA) has emerged as a research focus as it allows multiple users to share the same resource block via successive interference cancellation (SIC), which greatly alleviates the spectrum scarcity \cite{bck}. There is a rich literature to validate the performance gain of the NOMA in both spectral efficiency and energy efficiency (EE) in various scenarios. Note that the performance of NOMA is more sensitive to the power allocation than OMA, which makes the power allocation scheme the central consideration of the NOMA. Traditional power allocation schemes mainly utilize the convex optimization (CVXopt)-based approaches. For instance, in \cite{gzhang} and \cite{Jliu}, the total age of information was minimized  and the EE was maximized, respectively, for simultaneous wireless information and power transfer enabled NOMA systems  via the successive convex approximation (SCA) method. In \cite{8016604},  a distributed matching algorithm aiming to optimize the user pairing and power allocation for downlink NOMA was proposed. Nevertheless, the CVXopt-based approaches can explore the performance limit but may not be computationally efficient due to involving iterations \cite{rhang}.

 Recently, the deep learning (DL), especially the graph neural network (GNN), has been regarded as a new paradigm to address optimization problems of wireless networks. The advantages of these learning methods are two-fold\cite{lugnn}. First, a well-trained GNN supports real-time computation with limited performance loss. Second, the permutation invariance and equivariance property ensures the GNN to be scalable to the input sizes (e.g., number of users). Therefore, the GNN has been recognized as a basic model to support the power allocation for time-varying and dynamic wireless networks. In the literature, there are various works on DL-enabled NOMA. For example, DL was adopted to realize  adaptive user pairing and sum-rate maximization for NOMA systems in  \cite{ggui} and \cite{jjcui}, respectively. We note that these existing works are based on traditional neural networks with limited feature extracting capability. In \cite{GNNNOMA}, GNN was employed to achieve communication-efficient distributed scheduling for a multi-cell cluster-free NOMA  network. In particular, the vanilla GNN was considered in \cite{GNNNOMA}, which may suffer from a large generalization loss particular in the case with large-scale problem sizes. Some existing works have validated the necessity of involving model enhancement mechanisms, such as the multi-head attention in the GNN. As seen in \cite{dl,dl2,dl3}, the use of graph attention networks (GAT) was superior to the vanilla GNN in terms of EE maximization, sum-rate maximization and max-min rate. But the aforementioned GATs were originally designed for OMA scenarios and hence, cannot be extended to the NOMA scenarios.

This paper proposes a GAT-based power allocation scheme for NOMA, termed NOMANet. We formulate a quality of service quality of service (QoS) constrained  EE maximization problem for a multiple-subchannel multiple-user NOMA network and address it by NOMANet. Particularly, NOMANet employs the multi-head attention mechanism to extract the inter-user interaction within one subchannel. Furthermore, three versions of NOMANet are developed based on plain update, residual connection, and dense connection \cite{Gli}. The output of  NOMANet can be feasible via the customized activation function and the penalty method.  NOMANet is reused  by all subchannels and trained unsupervised. Numerical results evaluate  NOMANet in terms of optimality, feasibility rate, scalability and inference time. Overall, NOMANet, especially with residual and dense connections, achieves close performance to the SCA method but with millisecond-level inference, and is scalable to both users and subchannels. Moreover, we use the ablation experiment to discuss the over-smoothing issue in NOMANet.

\section{System Model and Problem Formulation}

Consider a downlink NOMA network, where a base station (BS) transmits signals to $K$ user equipments (UEs) using $N$ subchannels. The bandwidth of each subchannel is $B$. The number of UEs assigned to the $n$-th subchannel is $K_n$ and  $\sum\nolimits_{i = 1}^N {{K_i}}  = K$.

The symbol transmitted on the $n$-th subchannel is given by
\begin{flalign}
{x_n} = \sum\nolimits_{i = 1}^{{K_n}} {\sqrt {{p_{i,n}}} {s_{i,n}}},
\end{flalign}
where ${s_{i,n}}$ denotes the modulated symbol for the $i$-th UE on the $n$-th subchannel and 
 $p_{i,n}$ denotes the corresponding power allocation.

 The received signal at the $i$-th UE on the $n$-th subchannel is given by
\begin{flalign}
{y_{i,n}} = \sqrt {{p_{i,n}}} {h_{i,n}}{s_{i,n}} + \sum\nolimits_{j \ne i}^{{K_n}} {\sqrt {{p_{j,n}}} {h_{i,n}}{s_{j,n}}}  + {z_{i,n}},
\end{flalign}
where ${h_{i,n}}$ denotes the $i$-th UE's channel gain on the $n$-th subchannel and ${z_{i,n}}\in{\cal CN}(0,\sigma_{i,n}^2)$ denotes the additive white Gaussian noise.

We define $H_{i,n} \triangleq {| {{h_{_{i,n}}}} |^2}/\sigma _{i,n}^2$. The adopted SIC decoding order is based on the increasing order of $\{H_{i,n}\}_i$. Suppose that the $K_n$ UEs are allocated with the order ${H_{1,n}} \ge \cdots \ge {H_{i,n}} \ge \cdots \ge {H_{{K_n},n}}$, and the power allocation satisfy ${p_{1,n}} \le \cdots \le {p_{i,n}} \le \cdots \le {p_{{K_n},n}}$. On the $n$-th subchannel, the $i$-th UE can successfully decode the symbols for $j$-th UE with $j>i$ and remove the corresponding interference. With the NOMA protocol, the signal-to-interference-plus-noise ratio (SINR) of the $i$-th UE  on the $n$-th subchannel is given by
\begin{flalign}
{\rm{SINR}}_{i,n} = \frac{{{p_{i,n}}{H_{i,n}}}}{{1 + \sum\nolimits_{j = 1}^{i - 1} {{p_{j,n}}{H_{i,n}}} }}.
\end{flalign}

The EE of the considered system is defined as
\begin{flalign}
{\rm EE}\left(\left\{ {{p_{i,n}}} \right\}\right)   = \frac{{\sum\nolimits_{n = 1}^N {\sum\nolimits_{i = 1}^{{K_n}} {{{\log }_2}\left( {1 + {\rm{SINR}}_{i,n}} \right)} } }}{{\sum\nolimits_{n = 1}^N {\sum\nolimits_{i = 1}^{{K_n}} {{p_{i,n}}} }  + {P_{\rm{C}}}}}.
\end{flalign}

Our goal is to maximize the system EE subject to the UE's individual data requirement and the power budget of the BS, which is expressed as the following optimization problem:
\begin{subequations}\label{p0}
\begin{align}
\mathop {\max} \limits_{\left\{{{p_{i,n}}} \right\}} ~&{\rm EE}\left(\left\{ {{p_{i,n}}} \right\}\right) \label{cons:p0:A}\\
{\rm s.t.}~& {\log _2}\left( {1 + {\rm{SINR}}_{i,n}} \right) \ge {R_{{\rm{Req}}}} ,\label{cons:p0:B}\\
&\sum\nolimits_{n = 1}^N \sum\nolimits_{i = 1}^{{K_n}} {{p_{i,n}}}   \le {P_{\rm Max}},\label{cons:p0:C}\\
&{p_{1,n}} \le \cdots \le {p_{i,n}} \le \cdots \le {p_{{K_n},n}},\label{cons:p0:D}
\end{align}
\end{subequations}
where ${R_{\rm Req}}$ denotes the minimal information rate requirement, and ${P_{\rm Max}}$ denotes the power budget.

\begin{figure}
  \centering
  \includegraphics[width=0.40\textwidth]{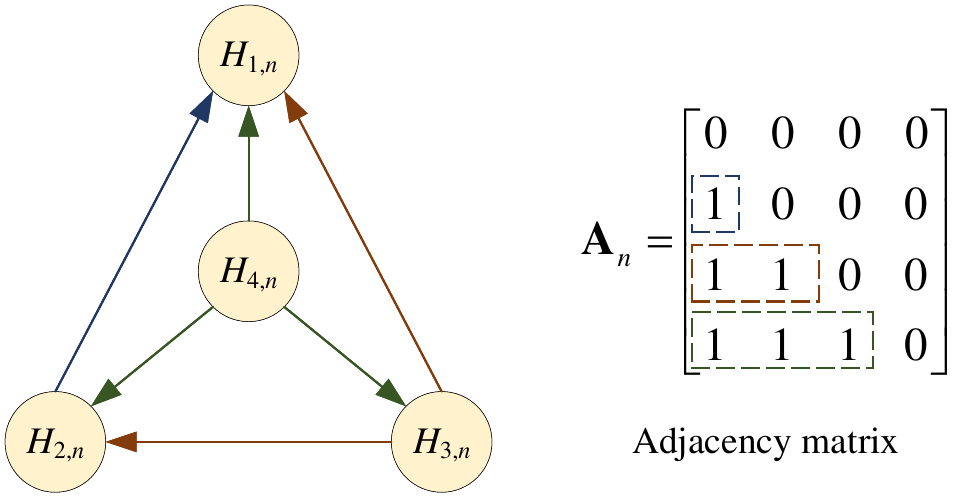} 
  \caption{Graph representation of the $n$-th subchannel: a four-UE case.}
  \label{gr}
\end{figure}

\section{NOMANet for Solving Problem \eqref{p0}}

\begin{figure*}[t]
  \centering
  \includegraphics[width=0.98\textwidth]{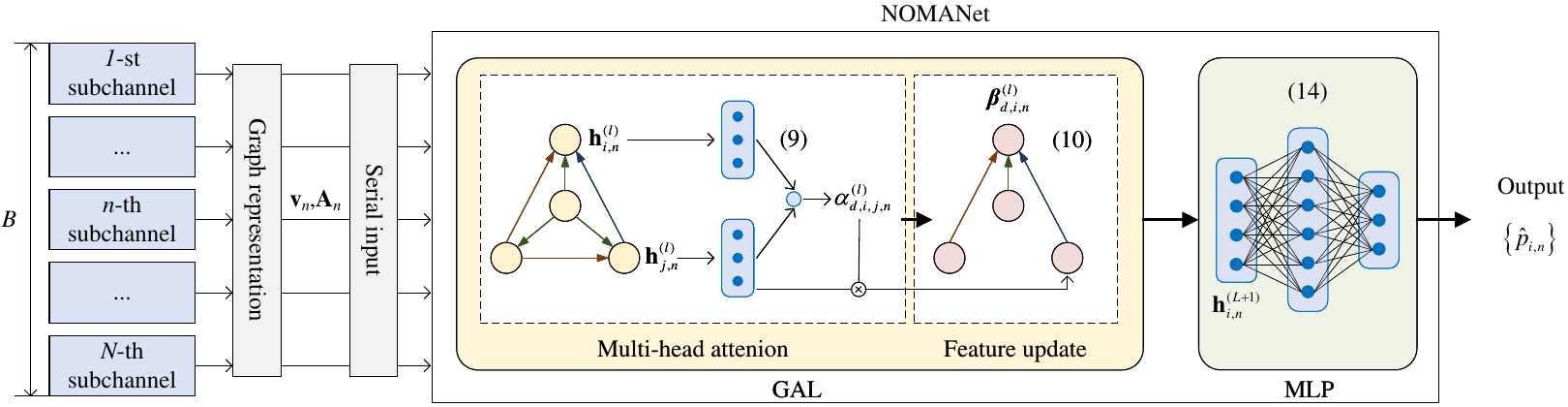}
  \caption{The structure of NOMANet.}
\end{figure*}

\begin{figure}[t]
  \centering
  \includegraphics[width=0.48\textwidth]{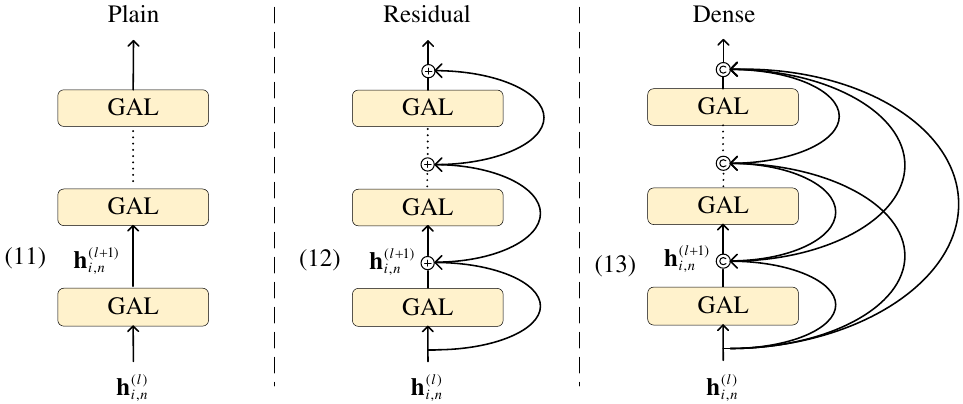}
  \caption{ The illustration of plain  update, residual connection and dense connection.}
\end{figure}

This section presents NOMANet for solving Problem \eqref{p0}. First, we represent  the NOMA system parameters into a graph structure, so that the relationship between the UEs sharing the same subchannel can be established. Then, the graph is input into  NOMANet to extract the hidden features among links associated with UEs, and these extracted features are further decoded as the desired power allocation. Finally, we adopt the unsupervised learning to train NOMANet to enhance the problem solving capability.

\subsection{Graph Representation of NOMA Network}

The considered NOMANet can be modeled as $N$ subgraphs, and each subgraph represents one subchannel as shown in Fig. \ref{gr}. The $n$-th subgraph is denoted by ${\cal G}_{n} = ({\cal V}_{n},{\cal E}_{n},{\bf A}_{n},{\bf v}_{n})$\footnote{In the proposed graph representation, there is no edge feature like \cite{dl}.}. Here, ${\cal V}_{n}$ and ${\cal E}_{n}$ represent the node set and the edge set, respectively, where each node represents one of the links sharing the subchannel and each edge represents the interaction between two corresponding links. ${\bf A}_{n}\in {\mathbb R}^{K_n \times K_n}$ denotes the adjacency matrix with 
\begin{flalign}
\left[{\bf A}_{n}\right]_{i,j} = \left\{ \begin{array}{l}
1,~{\rm if} \ i > j\\
0,~{\rm if} \ i \leq j
\end{array} \right..
\end{flalign}
${\bf v}_n\triangleq[H_{1,n},...,H_{K_n,n}]^T \in {\mathbb R}^{K_n}$ denotes the node feature vector. 

 For the $n$-th subgraph, the desired mapping $\Pi_\theta$ can be given by
\begin{flalign}\label{mapping}
     {\bf v}_n,{\bf A}_n\xrightarrow[]{\Pi_{\bm\theta}\left(\cdot\right)}\left\{\widehat{p}_{i,n}\right\}_i,
\end{flalign}
where $\bm\theta$ denotes the learnable parameter, and $\{\widehat{p}_{i,n}\}_i$ are the output of the GNN. To guarantee that $\{\widehat{p}_{i,n}\}$ satisfy the overall power constraint, i.e., \eqref{cons:p0:C}, the following activation function is adopted 
\begin{flalign}
{{{{\widehat p}_{i,n}}}:= {\frac{P_{{\rm{max}}}}{{{\rm{max}}\left( {{P_{{\rm{max}}}},\sum\limits_{n = 1}^N \sum\limits_{i = 1}^{{K_n}} {{\rm{ReLU}}\left( {{{\widehat p}_{i,n}}} \right)} } \right)}}} {\rm{ReLU}}\left( {{{\widehat p}_{i,n}}} \right).}
\end{flalign}

\subsection{Structure of NOMANet}

The proposed NOMANet includes two key components, i.e., $L$  graph attention layers (GALs) and MLP. Particularly, the GALs are used to extract the hidden features from the subgraphs, and the MLP is to map the extracted features into the desired power allocation. The detailed processes of the two components for the $n$-th subgraph are given as follows.

\subsubsection{Graph attention layer} Denote the input and output node features of the $l$-th GAL\footnote{For the $1$-st GAL, the input node features are $\{H_{i,n}\}_i$ and $F(1)$=1.} by $\{{\bf h}_{i,n}^{(l)}\in {\mathbb R}^{F(l)}\}_i$ and $\{{{\bf h}}_{i,n}^{(l+1)} \in {\mathbb R}^{F(l+1)}\}_i$, where $F(l)$ and $F(l+1)$ denote the corresponding feature dimensions\footnote{Note that $F(l+1)$ may vary due to different update strategies.}, respectively. The GAL consists of three operations, namely, multi-head attention, message aggregation and feature update.

The multi-head attention is to assign different weight scores to the neighboring nodes of one node. For the $i$-th node, the adjacency matrix defines its neighboring node set denoted by ${\cal N}(i)$. Assume that $D(l)$ attention heads are adopted. For the $d$-th attention head, the attention coefficient associated with the $j$-th node ($j\in {\cal N}(i)$) is given by
\begin{flalign}
\alpha_{d,i,j,n}^{(l)} = \frac{\exp(f_{1}({\bf W}_{d,1}^{(l)}{\bf h}_{i,n}^{(l)}+{\bf W}_{d,2}^{(l)}{\bf h}_{j,n}^{(l)})^{T}{\bf a}_{d}^{(l)})}{\sum_{k \in {\mathcal N}(i)}\exp\left(f_{1}({\bf W}_{d,1}^{(l)}{\bf h}_{i,n}^{(l)}+{\bf W}_{d,2}^{(l)}{\bf h}_{k,n}^{(l)})^{T}{\bf a}_{d}^{(l)}\right)},
\end{flalign} 
where ${\bf a}_{d}^{(l)} \in \mathbb {R}^{{{F}(l+1)/D(l)}}$  ,${\bf W}_{d,1}^{(l)} \in \mathbb{R}^{{\left({F}(l+1)/D(l)\right)\times{F}(l)}}$ and ${\bf W}_{d,2}^{(l)} \in \mathbb{R}^{{\left({F}(l+1)/D(l)\right) \times{F}(l)}}$ are the learnable parameters, and $f_{1}(\cdot)$ represents a non-linear functional operation.

Then, the $i$-th node aggregates its neighboring node features using the attention coefficients of the $d$-th attention head as 
\begin{flalign}
  {\bm \beta}_{d,i,n}^{(l)} = f_{2}\left(\alpha_{d,i,k,n}^{(l)}{\bf W}_{d,2}^{(l)}{\bf h}_{k,n}^{(l)}\vert k\in{\mathcal{N}(i)}\right)\in {\mathbb R}^{{F}(l+1)/D(l)},
\end{flalign}
where $f_{2}(\cdot)$ represents a permutation-invariance function.

After obtaining $\{{\bm \beta}_{d,i,n}^{(l)}\}_d$, three node feature update functions are considered, namely, plain update, residual connection and dense connection, as follows:
\begin{flalign}
{\bf h}_{i,n}^{(l+1)} &= \textup{Concat}\left(\{{\bm\beta}_{d,i,n}^{(l)}\}_{d=1}^{D(l)}\right) \in {{\mathbb{R}^{F(l+1)}}},\label{eq:plain}\\
   {\bf h}_{i,n}^{(l+1)} & = \textup{Concat}\left(\{{\bm\beta}_{d,i,n}^{(l)}\}_{d=1}^{D(l)}\right) + {\bf W}^{(l)}_{\rm  res} {\bf h}_{i,n}^{(l)} \in {{\mathbb{R}^{F(l+1)}}},\label{eq:res}\\
   {\bf h}_{i,n}^{(l+1)} & = \textup{Concat}\left(\textup{Concat}\left(\{{\bm\beta}_{d,i,n}^{(l)}\}_{d=1}^{D(l)}\right), {\bf h}_{i,n}^{(l)} \right)\label{eq:den}\\
   &= \textup{Concat}\left(\textup{Concat}\left(\{{\bm\beta}_{d,i,n}^{(l)}\}_{d=1}^{D(l)}\right),..., {\bf h}_{i,n}^{(1)},H_{i,n} \right)\nonumber\\
   &\in {{\mathbb{R}^{\sum_{i=0}^{l+1} F(i)}}},\nonumber
\end{flalign}
where $\textup{Concat($\cdot$)}$ represents the concatenation operation, and ${\bf W}^{(l)}_{\rm  res}\in{\mathbb R}^{F(l+1)\times F(l)}$ denotes the learnable parameters for the residual connection.

\subsubsection{MLP} The MLP is used to map the  extracted node feature to the desired power allocation for each node through  a node-level readout approach.   

For the $i$-th node, the input of the MLP is ${\bf h}_{i,n}^{(L+1)}$ while the corresponding output is ${\widehat p}_{i,n}$, which is given by
\begin{flalign}
    {\widehat p}_{i,n} = {\rm MLP}\left({\bf h}_{i,n}^{(L+1)}\left|{\bf W}_{3}\right. \right)\in \mathbb{R},
\end{flalign}
where ${\bf W}_{3}$ denotes the corresponding learnable parameters due to the linear transformation, whose dimensionality is independent of  $K_n$ as it is applied to each node.

As  observed, all learnable parameters in  NOMANet are independent of the number of UEs, i.e., $K_n$, and the number of subchannels, i.e., $N$, due to parameter sharing. Therefore, NOMANet is applicable to various input sizes. 

\subsection{NOMANet Training}

As mentioned in Section III-A, each subchannel is represented by a subgraph, which leads to $N$ subgraphs with different numbers of nodes (i.e., UEs). NOMANet is reused for all subgraphs as it is acceptable to different sizes of subgraphs. Particularly, the subgraphs are serially input into NOMANet to obtain $\{\Pi_{\bm \theta}({\bf v}_n,{\bf A}_n)\}_n$, i.e., $\{{\widehat p}_{i,n}\}_{i,n}$ (cf. \eqref{mapping}), which are further input into the loss function.

To alleviate the overhead of collecting labeled dataset, the unsupervised learning is adopted to train NOMANet. Besides, the penalty method is adopted to satisfy the constraints \eqref{cons:p0:B} and \eqref{cons:p0:D}. Then, the loss function is given by
\begin{flalign}
{\cal L}_{\bm\theta} =& \frac{1}{{\rm EE}\left(\left\{ {{{\widehat p}_{i,n}}} \right\}\right)} + \nonumber\\
&\lambda_1 \sum\nolimits_{n=1}^{N}\sum\nolimits_{i=1}^{K_n}{\rm ReLu}\left( {R_{{\rm{Req}}}} -  {\log _2}\left( {1 + {\rm{SINR}}_{i,n}} \right) \right) + \nonumber\\
&\lambda_2 \sum\nolimits_{n=1}^{N}\sum\nolimits_{i=2}^{K_n}{\rm ReLu}\left( {\widehat p}_{i-1,n} - {\widehat p}_{i,n}  \right),\label{loss}
\end{flalign}
where $\lambda_1$ and $\lambda_2$ are hyperparameters which represent the penalty weights due to the constraint violation.

\section{Numerical Results}

This section provides numerical results to evaluate the proposed NOMANet in terms of generalization performance.

\subsection{Simulation Setting}
\subsubsection{Simulation scenario} 
The system parameters are set by $N\in\{8,10,12\}$, $K_n=K \in \left\{4,5,6\right\}$, $\forall n\in {\cal N}$,  ${P_{\rm max}}=10$ W, ${P_{\rm C}}= 1$ W, ${R_{\rm Req}}= 0.1$ bit/s/Hz. The channel gains $\{ {h}_{i,n}\in{\mathbb C}\}$ are based on the Rayleigh distribution with average signal-to-noise-ratio being $20$ dB for both training samples and test samples. The labels for the test samples are obtained by traditional CVXopt-based algorithms. Besides, we use $N_{\rm Tr}$/$K_{\rm Tr}$ and $N_{\rm Te}$/$K_{\rm Te}$  to respectively denote the values of $N$/$K$ in the training phase and test phase, where $N_{\rm Te}$/$K_{\rm Te}$ could be different from $N_{\rm Tr}$/$K_{\rm Tr}$ during the test phase.

\begin{table}[t]
    \centering
    \caption{Datasets.}
    \label{Datasets}
    \begin{tabular}{c|c|c|c|c}
    \hline
    No. &$N$ &$K$ & Size& Type\\
     \hline
      \hline
     1   &10&5& 10,000& A\\
     \hline
     2   &10&4& 1,000& B\\
     \hline
     3   &10&5& 1,000& B\\
     \hline
     4   &10&6& 1,000& B\\
     \hline
     5   &8&4& 1,000& B\\
     \hline
     6   &8&5& 1,000& B\\
     \hline
     7   &8&6& 1,000& B\\
     \hline
     8   &12&4& 1,000& B\\
     \hline
     9   &12&5& 1,000& B\\
     \hline
     10   &12&6& 1,000& B\\
     \hline
    \end{tabular}
     \begin{tablenotes}
            \footnotesize
            \item {Type A: The training set, validation set and test set are all included.}
            \item {Type B: Only test set is included.}
    \end{tablenotes}
\end{table}

\begin{table*}[t]
\centering
\caption{Performance evaluation: $(N_{\rm Tr},K_{\rm Tr})=(10,5)$.}\label{result}
\begin{tabular}{c|c||c|c|c|c|c|c|c|c|c|c}
\hline
  \multirow{3}{*}{$N_{\rm Te}$} &  \multirow{3}{*}{$K_{\rm Te}$} & \multicolumn{2}{c|}{\multirow{2}{*}{CVX}} & \multicolumn{2}{c|}{\multirow{2}{*}{MLP}}  & \multicolumn{6}{c}{NOMANet}\\
\cline{7-12}
 & & \multicolumn{2}{c|}{} & \multicolumn{2}{c|}{} & \multicolumn{2}{c|}{GAT-Plain} & \multicolumn{2}{c|}{GAT-Res} & \multicolumn{2}{c}{GAT-Dense}\\
\cline{3-12}
 &  & OP & FR & OP & FR & OP/SC & FR & OP/SC & FR & OP/SC & FR\\
 \hline
 \hline

\cellcolor{white}~&\cellcolor{white}4&{$\times$}&{$\times$}&{$\times$}&{$\times$}&{85.27\%}$^\S$&83.60\%&{98.42\%}$^\S$&{94.20\%}&{98.70\%}$^\S$&{88.60\%}\\

\cellcolor{white}~8&5&{$\times$}&{$\times$}&{$\times$}&{$\times$}&{80.23\%}$^\dag$&{78.10\%}&{97.44\%}$^\dag$&{91.60\%}&{97.70\%}$^\dag$&{87.40\%}\\

\cellcolor{white}~&\cellcolor{white}6&{$\times$}&{$\times$}&{$\times$}&{$\times$}&{71.39\%}$^\S$&{70.50\%}&{95.94\%}$^\S$&{90.40\%}&{95.40\%}$^\S$&{80.80\%}\\
 \rowcolor{blue!10} 
&4&{$\times$}&{$\times$}&{$\times$}&{$\times$}&{82.44\%}$^\P$&{80.00\%}&{98.73\%}$^\P$&{93.60\%}&{98.46\%}$^\P$&{86.60\%}\\

 \rowcolor{blue!10} 
10&5&{100\%}&{100\%}&{64.27\%}&{63.80\%}&{88.20\%}&{79.20\%}&{97.36\%}&{90.80\%}&{97.52\%}&{85.20\%}\\

 \rowcolor{blue!10} 
&6&{$\times$}&{$\times$}&{$\times$}&{$\times$}&{81.42\%}$^\P$&{70.40\%}&{98.26\%}$^\P$&{86.80\%}&{98.05\%}$^\P$&{76.00\%}\\

 \rowcolor{orange!10} 
&4&{$\times$}&{$\times$}&{$\times$}&{$\times$}&{77.28\%}$^\S$&{81.60\%}&{97.68\%}$^\S$&{90.40\%}&{97.06\%}$^\S$&{83.00\%}\\
 
 \rowcolor{orange!10} 
12&5&{$\times$}&{$\times$}&{$\times$}&{$\times$}&{79.60\%}$^\dag$&{72.20\%}&{97.66\%}$^\dag$&{90.60\%}&{97.89\%}$^\dag$&{86.40\%}\\
\rowcolor{orange!10} 
&6&{$\times$}&{$\times$}&{ $\times$}&{$\times$}&{71.45\%}$^\S$&{68.80\%}&{94.30\%}$^\S$&{89.60\%}&{97.36\%}$^\S$&{82.60\%}\\
 \hline
 \multicolumn{2}{c||}{Inference time}& \multicolumn{2}{c|}{2.8 s} &\multicolumn{2}{c|}{1.74 ms} &\multicolumn{2}{c|}{3.72 ms} & \multicolumn{2}{c|}{3.94 ms} & \multicolumn{2}{c}{4.39 ms} \\
 \hline
\end{tabular}
\begin{tablenotes}
        \footnotesize
        \item
        $\times$ represents ``not applicable".
        \item{$^\dag$/$^\P$/$^\S$: The result marked with $^\dag$/$^\P$/$^\S$ represents the scalability performance with $N_{\rm Te} \ne N_{\rm Tr}$/$K_{\rm Te} \ne K_{\rm Tr}$/$N_{\rm Te} \ne N_{\rm Tr}$ and $K_{\rm Te} \ne K_{\rm Tr}$.}
        \item OP/SC/FR: Optimality performance/scalability performance/feasibility rate.
\end{tablenotes}
\end{table*}

\subsubsection{Computer configuration} All DL models are trained and tested by Python 3.9.13 with Pytorch 1.13.1 on a computer with Intel 12600KF CPU and NVIDIA RTX 3080 (10 GB of memory).

\subsubsection{Initialization and training} The learnable parameters are initialized according to He (Kaiming) method and the learning rate is initialized as $10^{-3}$. The $optim$ algorithm is adopted as the optimizer during the training phase. The batch size is set to $16$ for $50$ training epochs. The learnable parameters with the best performance are used as the training results.

\subsubsection{Baselines}
 NOMANets with the plain update, residual connection and dense connection are termed ``GAT-Plain", ``GAT-Res" and ``GAT-Dense", respectively. To numerically evaluate  NOMANets, the following baselines are considered, i.e.,
\begin{itemize}
\item {CVXopt-based approach}:  An SCA-based optimization algorithm, similar to \cite{lusee}, termed  ``CVX". 
\item {MLP}:  A classical feed-forward  neural network, similar to \cite{mlp}, termed  ``MLP". 
\end{itemize}

\subsubsection{Test performance metrics} The following performance metrics are adopted for the evaluation. 
\begin{itemize}
  \item 
  Optimality performance: The ratio of the average achievable EE by the DL model to the EE obtained by the CVXopt-based approach with $N_{\rm Te}=N_{\rm Tr}$ and $K_{\rm Te} = K_{\rm Tr}$.
  \item 
  Feasibility rate: The percentage of the feasible solutions to the considered problem.
  \item 
  Scalability performance: The ratio of the average achievable EE by the DL model to the EE obtained by the CVXopt-based approach with with $N_{\rm Te}\ne N_{\rm Tr}$ or/and $K_{\rm Te} \ne  K_{\rm Tr}$. 
  \item
  Inference time per subchannel: Average running time to yield the feasible power allocation by the DL model for one subchannel.
\end{itemize}

\subsection{Result Analysis}

Table  \ref{result} evaluates  NOMANets  in terms of the performance metrics, and we have the following observations.  
\subsubsection{Optimality and Inference time} For $N_{\rm Tr}=N_{\rm Te}$ and $K_{\rm Tr}=K_{\rm Te}$,  NOMANets, especially GAT-Res and GAT-Dense, achieve comparable EE to the traditional CVXopt-based approach, but use much smaller inference time. Besides,  NOMANets outperform the MLP thanks to the multi-head attention mechanism, while the GAT-Res and the GAT-Dense are  superior to the GAT-Plain as the expressive capability is enhanced by mitigating the over-smoothing issue via residual/dense connection.
\subsubsection{Scalability and Feasibility rate} For $N_{\rm Tr}\ne N_{\rm Te}$ or/and $K_{\rm Tr}\ne K_{\rm Te}$, NOMANets are still applicable due to the parameter sharing among nodes. That is, NOMANets are applicable to the cases with various values of $N$ or/and $K$ without re-training and capable of handling the problem sizes unseen during the training phase. It is observed that the GAT-Res and the GAT-Dense experience limited performance fluctuation when scalable to unseen problem sizes, which is more stable than the GAT-Plain. The reason can be found in \eqref{eq:plain}-\eqref{eq:den}, where it is observed that the plain updating is a special case of the residual/dense connection, which indicates that the GAT-Res/Dense is superior to the GAT-Plain with enough training samples. There are still some space to improve for the feasibility rate of NOMANets, specifically for scalability. The main reason is that it is challenging for the penalty method  to guarantee $(K\times N)$ QoS constraints to be satisfied. In general, the GAT-Res yields a higher feasibility rate than the GAT-Dense, as too many connections may also struggle the training phase.

\subsection{Ablation experiment}

Table \ref{table:ablation_rd} presents the results of the ablation experiment to demonstrate the effectiveness of residual or dense connections against the over-smoothing issue. As seen, the optimality performance of NOMANets first increases and then decreases with the depth, but the performance degradation is mitigated by the residual and dense connections at the expense of a small increase in inference time (cf. Table \ref{result}). 

\begin{table}[t]
\centering
\caption{Ablation experiment: $(N_{\rm Te},K_{\rm Te}) = (N_{\rm Tr},K_{\rm Tr})=(10,5)$.}
\begin{tabular}{c||c|c|c|c}
\hline

{\diagbox [width=6em] {Model}{OP}{Depth}}   &1  & 2 & 3 & 4 \\

\hline
\hline
{GAT-Plain} &  85.88\% &  88.20\% &  85.90\% & 84.01\%    \\
\hline
{GAT-Res}  &  95.01\% &  97.36\% &  96.51\% & 96.26\%    \\
\hline
{GAT-Dense}  &  92.32\% &  97.52\% &  97.13\% & 97.02\%    \\
\hline

\end{tabular}
\label{table:ablation_rd}
 \begin{tablenotes}
        \footnotesize
       \item Depth: Number of GALs.
\end{tablenotes}
\end{table}

\section{Conclusion}
This paper has proposed a GNN enabled power allocation scheme for NOMA networks, named NOMANet. An EE maximization problem has been formulated under constraints of the individual rate requirement, power budget and NOMA protocol and then, addressed by NOMANet. To enhance the expressive capability, NOMANet employs the multi-head attention and residual/dense connection. Besides, NOMANet has been trained unsupervised with learnable parameters independent of the numbers of UEs and subchannels. Numerical results have demonstrated that NOMANet achieved close performance to the SCA-based algorithm but much more computationally efficient. The scalability of the NOMANet to both UEs and subchannels was also validated.


\begin{thebibliography}{10}
\bibitem{bck}
Z. Ding et al., ``Application of non-orthogonal multiple access in LTE and 5G networks," \emph{IEEE Commun. Mag.}, vol. 55, no. 2, pp. 185-191, Feb. 2017.

\bibitem{gzhang}
G. Zhang, Y. Lu, Y. Lin, Z. Zhong, Z. Ding, and D. Niyato, ``AoI minimization in RIS-aided SWIPT systems," \emph{IEEE Trans. Veh. Technol.}, vol. 73, no. 2, pp. 2895-2900, Feb. 2024.

\bibitem{Jliu}
J. Liu, K. Xiong, Y. Lu, P. Fan, Z. Zhong, and K. B. Letaief,  ``SWIPT-enabled full-duplex NOMA networks with full and partial CSI,"  \emph{IEEE Trans. Green Commun. Networking}, vol. 4, no. 3, pp. 804-818, Sept. 2020.

\bibitem{8016604}
W. Liang, Z. Ding, Y. Li, and L. Song. ``User pairing for downlink non-orthogonal multiple access networks using matching algorithm," \emph{IEEE Trans. Commun.}, vol. 65, no. 12, pp. 5319-5332, Dec. 2017.

\bibitem{rhang}
R. Zhang, K. Xiong, Y. Lu, D. W. K. Ng, P. Fan, and K. B. Letaief, ``SWIPT-Enabled cell-free massive MIMO-NOMA networks: A machine learning-based approach," \emph{IEEE Trans. Wireless Commun.}, vol. 23, no. 7, pp. 6701-6718, July 2024.

\bibitem{lugnn}
Y. Lu, Y. Li, R. Zhang, W. Chen, B. Ai, and D. Niyato, ``Graph neural networks for wireless networks: Graph representation, architecture and evaluation," early accessed in \emph{IEEE Wireless Commun.}, 2024.

\bibitem{ggui}
R. H. Y. Perdana, T.-V. Nguyen, and B. An, ``Adaptive user pairing in multi-IRS-aided massive MIMO-NOMA networks: Spectral efficiency maximization and deep learning design,"  \emph{IEEE Trans. Commun.}, vol. 71, no. 7, pp. 4377-4390, July 2023.

\bibitem{jjcui}
H. Huang, Y. Yang, Z. Ding, H. Wang, H. Sari, and F. Adachi, ``Deep learning-based sum data rate and energy efficiency optimization for MIMO-NOMA systems,"  \emph{IEEE Trans. Wireless Commun.}, vol. 19, no. 8, pp. 5373-5388, Aug. 2020.


\bibitem{GNNNOMA}
X. Xu, Y. Liu, Q. Chen, X. Mu, and Z. Ding, ``Distributed auto-learning GNN for multi-cell cluster-free NOMA communications,"  \emph{IEEE J. Sel. Areas Commun.}, vol. 41, no. 4, pp. 1243-1258, April 2023.

\bibitem{dl}
Y. Li, Y. Lu, R. Zhang, B. Ai, and Z. Zhong, ``Deep learning for energy efficient beamforming in MU-MISO networks: A GAT-based approach," \emph{IEEE Wireless Commun. Lett.}, vol. 12, no. 7, pp. 1264-1268, Jul. 2023.

\bibitem{dl2}
Y. Li, Y. Lu, B. Ai, O. A. Dobre, Z. Ding, and D. Niyato, ``GNN-based beamforming for sum-rate maximization in MU-MISO networks," \emph{IEEE Trans. Wireless Commun.}, vol. 23, no. 8, pp. 9251-9264, Aug. 2024.

\bibitem{dl3}
Y. Li, Y. Lu, B. Ai, Z. Zhong, D. Niyato, and Z. Ding, ``GNN-enabled max-min fair beamforming," \emph{IEEE Trans. Veh. Technol.}, vol. 73, no. 8, pp. 12184-12188, Aug. 2024.



\bibitem{Gli}
G. Li et al., ``DeepGCNs: Making GCNs go as deep as CNNs," \emph{IEEE Trans. Pattern Anal. Mach. Intell.}, vol. 45, no. 6, pp. 6923-6939, June 2023

\bibitem{lusee}
Y. Lu, ``Secrecy energy efficiency in RIS-assisted networks," \emph{IEEE Trans. Veh. Technol.}, vol. 72, no. 9, pp. 12419-12424, Sept. 2023.

\bibitem{mlp}
C. Hu et al., ``AI-empowered RIS-assisted networks: CV-enabled RIS selection and DNN-enabled transmission,"  \emph{IEEE Trans. Veh. Technol}, vol. 73, no. 11, pp. 17854-17858, Nov. 2024.
\end{thebibliography}
\end{document}